\UseRawInputEncoding
\documentclass[pra,12pt,tightenlines,nofootinbib]{revtex4} 
\pdfoutput=1
\usepackage{amsmath}
\usepackage{graphics, graphicx}
\usepackage{epsfig}
\usepackage{ulem}
\usepackage{color}

\begin{document}

\title{Generalized Quantum Mechanics}
\author{James  Hartle}
\affiliation{Department of Physics, University of California, Santa Barbara, California, 93106, USA} {\affiliation{ Santa Fe Institute, \\ 1399 Hyde Park Road,  Santa Fe, New Mexico  87501, USA.}  
\bibliographystyle{unsrt}

\begin{abstract}{\it A unified  framework framework for different formulations of quantum theory is introduced specifying  what is  meant by a quantum theory in general. }
\end{abstract}

\maketitle

\section{Introduction}
\label{intro}
It used to be that the term `quantum mechanics' referred to one thing --- the non-relativistic  (Copenhagen)  quantum mechanics of laboratory measurement situations.   There was only one formulation of quantum  theory in terms of a posited classical spacetime, a Hilbert space of states represented by wave functions on that spacetime, a Hamiltonian operator for dynamics, and operators representing different alternatives that might be measured. There was one formulation of quantum mechanics  that we can call {\it Copenhagen quantum mechanics (CQM). }

But, as our observations of the physical world  expanded, different formulations of quantum mechanics were required for different the applications of the theory.  Besides a formulation for measurement situations we now needed  formulations  for closed systems like the Universe that are that are not measured from the outside and are suitable  for quantum cosmology.   We needed formulations for quantum spacetime geometry, formulations for alternative theories like Bohm theory, formulations for predicting the future, formulations for retrodicting the past and formulations for measurement situations  that are more general than CQM such as   decoherent histories quantum theory(DH), consistent histories quantum theory, etc. \cite{HarJer,Harhouch} that apply to the quantum  Universe as a whole. 

The question  naturally arises:  What qualifies as a formulation of quantum theory?
 What do we mean by quantum mechanics generally?   What kind of theory qualifies as a formulation of quantum theory?  The answer proposed here is a general framework for all the different formulations quantum theory ---  {\it  generalized quantum mechanics }(GQM)\footnote{Other answers to this question have been proposed e.g, \cite{UnivQM}.} 
 
 GQM was first discussed in  \cite{Harhouch}. We offer a brief summary of it  in this paper.

\eject

 \section{The Three Ingredients of  A Generalized Quantum Theory} 
 \label{threeingred}
 
 To illustrate how generalized quantum theories are constructed we consider the example of a closed system like the interior of a large box with a fixed flat spacetime specified by theories of its quantum state  and quantum dynamics such as the no-boundary quantum state  \cite{NBWF,WINBWF} and a theory of quantum  gravity and matter fields inside.

A generalized quantum theory is specified by three ingredients:

\begin{itemize}  
 
\item{ 1 \it A  set of possible fine-grained histories:} These are the elements of a set of four-dimensional time-histories of spacetime metric and matter field configurations. They are one of the the most refined descriptions of the Universe it is possible to give.  For example the fine-grained histories  for a single particle are its possible single valued paths in spacetime,  field configurations in field theory,  and for gravity the possible manifolds with Lorentz signatured  geometries  in relativistic theories of gravitation, 

\item{ 2 \it Coarse-grained histories:} Partitions of the set of  fine-grained histories into an exhaustive set of exclusive  four-dimensional diffeomorphism invariant classes  $\{c_{\alpha}\}, \alpha=1,2 \cdots$ defines sets of coarse-grained histories. Sets of coarse-grained histories are the most general notion of alternative describable in spacetime terms for which a formulation of quantum theory predicts testable probabilities based on theories  of a closed Universe's quantum state and dynamics. 

\item {3 \it A measure of quantum interference between pairs of coarse-grained  histories
$(c_\alpha, c_{\alpha'})$ called a Decoherence Functional, $D(\alpha,\alpha')$ }  \label{Dfnal}  and incorporating theories of the Universe's initial state and quantum dynamics.   

Sets of alternative coarse-grained histories with negligible interference between all pairs of members are said to decohere, or to be consistent. It is then consistent to assign probabilities in an exhaustive set of alternative histories when, and only when, that set is decoherent. It is the criterion of decoherence, rather than any notion of measurement, which determines the consistency of the various formulations  of  quantum theory. 

A decoherence functional should satisfy the following four conditions:
\end{itemize}

The essential features of quantum mechanics are captured by any complex valued decoherence functional $D(\alpha,\alpha')$ that satisfies the following conditions \cite{Harhouch,Ish93}:

\renewcommand{\theenumi}{\roman{enumi}}
\begin{enumerate}

\item{} Hermiticity:  $D(\alpha,\alpha') = D^*(\alpha',\alpha)$,

\item{} Normalization:  $\sum_{\alpha\alpha'} D(\alpha,\alpha') =1 $, 
 
\item{} Positivity:  $D(\alpha,\alpha) \ge 0$,

\item{} Consistency with the Principle of Superposition: 

 This means the following:   Partitioning a set of histories $\{C_\alpha\}$ into coarser grained  sets $\{C_{\bar\alpha}\}$ is an operation of coarse graining. Every history $C_\alpha$ is in one  and only one of the sets $C_{\bar\alpha}$  a fact that we indicate schematically by $\alpha\in \bar\alpha$. 
This defines  a new set of alternative histories $\{C_{\bar\alpha}\}$ with elements
 \begin{equation}
 C_{\bar\alpha} = \sum_{{\alpha} \in {\bar\alpha}}  C_\alpha.
 \end{equation}
 related to the finer grained set by superposition:
 \begin{equation}
 D(\bar\alpha,\bar\alpha') = \sum_{\alpha\in\alpha}]\sum_{\alpha'\in\bar\alpha'}D(\alpha,\alpha').
 \end{equation}
 \end{enumerate}
 
 
 
                                                                                                                                                                                                                            


\section{Predicting Probabilities for Time Histories} 
\label{pred}
\centerline{\bf The central formula predicting the probabilities for}
 \centerline{\bf which histories happen in the Universe is then this:}
 
\begin{equation}
\label{dp}
\centerline{\fbox{$D(\alpha,\alpha')\approx\delta_{\alpha\alpha'}p(\alpha)$}}  
\end{equation}}

\noindent In \eqref{dp} 
Interference between all pairs of histories vanishes when the decoherence functional is diagonal.  The diagonal elements are the probabilities of the individual histories in a decoherent set. These probabilities satisfy all the usual rules of probability theory as a consequence of i-iv. The set of histories is then said to be `decoherent' or `consistent'. 
Usual Copenhagen quantum mechanics can be seen to  be a consequence of  i)-iv), see. e.g. \cite{GH90}.  

We now consider a few examples:

\section{The Hamiltonian Quantum Mechanics of a Non-Relativistic Particle as an Example of Generalized Quantum Theory}
\label{hamqm}
Consider a non-relativistic particle moving in one dimension $x$ of a flat classical spacetime. Suppose again that $x$ is divided up into a set of intervals 
$\{\Delta_k\}$,   $k=1,2, ...$   A coarse-grained description of  the particle's location at one time  would specify which interval  it is in at that time  but not where it is in that interval.  A coarse-grained history of the particle's motion might describe which intervals it occupies at a series of times, say,  $t_1,t_2,  \cdots t_n$. 

In quantum mechanics such alternative positions at a moment of time to $t$ are described by (Heisenberg picture) projection operators $P_k (t)$  that project onto the ranges $\Delta_k$ at time time $t$. A history $(\alpha_n,\alpha_{n-1}, \cdots $ where the particle is in various intervals $\alpha=(\alpha_n, \alpha_{n-1}, \cdots ,\alpha_1)$  at  times $t_1,t_2, \cdots t_n$  is  then described by the history operator
\begin{equation}
\label{hist}
C_\alpha\equiv P_{\alpha_n}(t_n)   \cdots P_2(t_2) P_1(t_1)   . 
\end{equation}

The decoherence functional   for this set of alternative histories is defined by:
\begin{equation}
D(\alpha,\alpha') \equiv Tr{(C_\alpha}^\dagger \rho C_{\alpha'}) .
\end{equation}
\vskip .2in
\noindent  which is  easily seen to satisfy properties (i)-(iv) above.  
Probabilities for betting on which history occurs  are then found using \eqref{dp}. 


\section {Sum-Over-Histories Quantum Mechanics for Theories with a Time as a Generalized Quantum Theory}
\label{sohqt} 

The fine-grained histories, coarse graining, and decoherence functional of a sum-over-histories quantum mechanics of a theory with a well defined physical time are specified as follows:

1. Fine-Grained Histories: The fine-grained histories are the possible paths in a configuration space of generalized coordinates {$q_i$} expressed as single-valued functions of the physical time $t$. Only one configuration is possible at each instant. Sum-over-histories quantum mechanics, therefore, starts from a unique fine-grained set of alternative histories of the Universe in contrast to Hamiltonian quantum mechanics that starts from many.

2. Allowed Coarse Grainings: There are many ways of partitioning the fine-grained paths into exhaustive and exclusive classes, {$c_\alpha$}. However, the existence of a physical time allows an especially natural coarse graining because paths cross a constant time surface in the extended configuration space $(t, q_i)$ once and only once. Specifying an exhaustive set of regions {$\Delta_k$}  $k =1,2,\cdots$ of the $q_i$  at one time partitions the paths into the class of those that pass through $\Delta_1$  at time $t_1$, the class of those that pass through $\Delta_2$ at time $t_2$, etc.  More generally, different exhaustive sets of regions {$\Delta_j $} at times ${t_j}$, $j = 1,ááá ,n$ similarly define a partition of the fine-grained histories into exhaustive and exclusive classes --- coarse-grained histories . More general partitions of the configuration space paths corresponding to alternatives that are not at definite moments of time are described in \cite{Harstcg}.

3. A Decoherence Functional: The decoherence functional for sum-over-histories quantum mechanics for theories with a well-defined time is.
\begin{equation}
\label{sohdf}
D(\alpha,\alpha')=\int_{\alpha}\delta q(t)  \int_{\alpha'}  \delta q'(t) \delta( q_f' -q_f)  \exp\{i(S[q'(t)]-S[q(t)]\}  
\rho(q'_0, q_0) .
\end{equation}

Here, we consider an interval of time from an initial instant $t = 0$ to some final time and $t = T$. The first integral is sum over paths $q'(t)$ that begin at q0, end at qf, and lie in the class $c_\alpha$ as the subscript $\alpha$  on the integral sign is meant imply. The integral includes an integration over q0 and qf . The second integral over paths q(t) is similarly defined. If $\rho (q',q)$ is a density matrix representing the assumed quantum state, then it is easy to verify that D defined by \eqref{sohdf} satisfies conditions (i) --- (iv) of Section III.

\section{Different Formulations of Quantum Theory}
\label{diff-form} Different notions of fine-grained histories, different notions of allowed coarse grainings, and and different mechanisms of decoherence will all lead to different formulations of quantum mechanics within the generalized quantum theory framework. In particular it may lead to  The different ways of formulating decoherent quantum theory \cite{HarDHrecords,Har08,HarGQMspacetime} Quantum mechanics is not just one thing it is many.


\section{The Challenge of Quantum Cosmology}
\label{qcosmo} The two simple examples of  generalized quantum theory in Sections IV and V  both assumed a fixed background classical spacetime. But in a quantum Universe  spacetime is not necessarily classical.   It's not classical near  a big bang or a big crunch  singularity for example.  In those situations  there will be quantum fluctuations away from classicality both large and small.
These deviations from classicality are important for understanding the large scale structure in the Universe that we see shortly after the big-bang in the cosmic background radiation and today in the large scale distribution of the galaxies.

In a quantum universe classical spacetime is not something to be posited but rather something to be calculated from the Universe's quantum state  like the no-boundary quantum state \cite{NBWF,WINBWF}   using a quantum formulation of cosmological spacetime geometry ---  a  quantum  theory  of gravity.  Even then classical spacetime will generally be available only in patches and exhibit quantum fluctuations away from classicality.

The construction of a generalized quantum mechanics for cosmological spacetimes requires the specification of the three ingredients that were the subject of Section II. Fine-grained histories would be the allowed set of  manifolds carrying a metric and matter fields.  Coarse grainings would be partitions of this set into 4d-diffeomorphism invariant classes. A decoherence functional would complete the specification. 

Candidates for these can all be specified in the context of linearized gravity.   But at the time of writing (autumn 2021) they are a challenge for future research (Although some progress has been made in lattice approximation to general relativity based on the Regge calculus 
 \cite{HarSimp1}). 

\section{Frameworks and Formulations of Physical Theory}
\label{framform} 
It is important to emphasize that generalized quantum mechanics is not a physical theory.  By itself it makes no predictions. Rather it is a framework incorporating general principles for unifying  formulations of quantum theory that do make predictions in different physical circumstances. 

There are a number of important theoretical structures  elsewhere in physics that have specific formulations in  a general framework of the kind  described this paper. Some examples follow:
  
One  instance  in quantum theory are the different formulations of decoherent histories quantum mechanics, e.g. \cite{HarDHrecords,Har08,Ohk93} under the same generalized quantum theory. 

Another example from quantum mechanics are the different decoherence functionals that  incorporate  different mechanisms of decoherence. With GQM one can construct quantum theories that incorporate, strong, medium, or  weak decoherence, and linear positivity. e.g. \cite{HarLP,Har08,HarGQMspacetime}. 

 Classical physics provides another example. The ingredients of a framework for classical physics would specify  the equations of motion, boundary conditions specifying solutions, and instructions for connecting those to realizable observations.   
 A classical theory is specified by a particular choice for these ingredients. The many choices produce very different theories. Think of the similarities and differences between  classical mechanics, hydrodynamics , electromagnetism, general relativity, turbulence, elasticity, etc, etc,

String theory is a general framework for different kinds of predictive string vacua.  It specifies observable predictions --- of classical spacetime and particular kinds of field theories in those spacetimes--- for example. 

The ubiquity of this  (general framework/specific formulation) structure  of many physical theories might be because  it is usually easier to work from specific examples to general principles than the other way around \cite{Harnatconf} 

\section{Conclusion}
\label{concl} 
A quantum theory of a closed physical system (like a spatially closed Universe) predicts quantum probabilities for what goes on inside the system by specifying notions of fine and coarse-grained histories together with a measure of interference between coarse-grained histories (a decoherence functional) incorporating theories of the system's quantum state and dynamics.
These probabilities can be used to bet on what will happen in the future, what  happened in the past, and what will be observed in any observational situation in the future. Thus, these probabilities enable science.

\vspace{.2in}

\noindent{\bf Acknowledgments:}  The work reported in this paper was supported over many years by the US National Science Foundation. The preparation of this   particular paper was supported by  NSF grant PHY-18-8018105. The author has benefited over the years with many discussions  with Mark Srednicki and the late Murray Gell-Mann.

\end{document}